# A single liquid chromatography procedure to concentrate, separate and collect size-selected polyynes produced by pulsed laser ablation in water

Sonia Peggiani, Anna Facibeni, Pietro Marabotti, Alessandro Vidale, Stefano Scotti & Carlo S. Casari

Taylor & Francis
Taylor & Francis Group

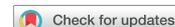

Check for updates

# A single liquid chromatography procedure to concentrate, separate and collect size-selected polyynes produced by pulsed laser ablation in water

Sonia Peggiani[a], Anna Facibeni[a], Pietro Marabotti[a], Alessandro Vidale[a], Stefano Scotti[b], and Carlo S. Casari[a]

[a]Micro- and Nanostructured Materials Laboratory, Department of Energy, Politecnico di Milano, Milan, Italy; [b]Shimadzu Italia Via GB Cassinis 7, Milan, Italy

**ABSTRACT**

Polyynes are linear carbon chains characterized by alternated single and triple bonds and terminated by hydrogen or other terminal substituents. They can be synthesized by pulsed laser ablation in liquid (PLAL) as a scalable, cost-effective, and fast physical technique. Water can be employed as a solvent for PLAL to avoid toxicity problems and to reduce costs compared to organic solvents. However, in in this case, the production yield of polyynes reached is extremely low and prevents further characterization and implementation in new functional materials. In this work, we synthesized polyynes by pulsed laser ablation in water and we optimized the process parameters to improve the yield of polyynes by PLAL. Then, we developed a procedure entirely based on reversed-phase high-performance liquid chromatography (RP-HPLC) which effectively enables the concentration, separation and collection of polyynes according to their length. Since the polyynes sample is an aqueous solution, we could inject it directly into the RP-HPLC column without the dilution step required in the case of a sample in an organic solvent. Thanks to our single RP-HPLC procedure, it is possible to highly increase the concentration and separately characterize different size-selected polyynes for further use in functional materials.

## 1. Introduction

Carbyne is the infinite one-dimensional wire made of sp hybridized carbon atoms, showing structure-dependent electronic properties and outstanding theoretically predicted properties (e.g., high thermal conduction, high electron mobility, high surface area, and high Young modulus).[1,2] Carbyne displays two isomeric configurations. One is the cumulene, a metal, composed of a sequence of consecutive CC double bonds and the other is the polyyne, a semiconductor, constituted by alternated single and triple bonds.[3] Polyynes are finite sp-carbon chains with chemicophysical properties strongly influenced by their length and termination. Well-known physical techniques such as submerged arc discharge in liquid (SADL) and pulsed laser ablation in liquid (PLAL) allow the synthesis of polyynes.[4–16] Besides the different organic solvents employed for the synthesis of polyynes, water has been also exploited due to its low cost, availability and non-toxicity, allowing the detection of hydrogen-capped polyynes ($C_nH_2$, where n indicates the number of carbon atoms).[17–19] To identify and separate sp-carbon chains depending on the length and termination,[20] we employed reversed-phase high-performance liquid chromatography (RP-HPLC). Having the polyynes in water media is an advantage of this technique in that the dilution step before sample injection, which would be necessary in the case of polyynes in organic solvent, is avoided. However, the synthesis in water implies a lower yield and poorer stability of polyynes compared to those obtained by laser ablation in organic solvents.[20–23] To perform further characterizations on size-selected polyynes and for the application of these structures in different technological fields, a high concentration of polyynes in an aqueous solution is needed.[24–26] Consequently, once the synthesis parameters are optimized to achieve the highest possible yield of polyynes by PLAL, post-synthesis concentration methods are also required.

The classic evaporative technique, such as vacuum rotary evaporation, is unsuitable for concentrating polyynes in aqueous solutions. Indeed, water has a too high boiling point that would require high temperatures of the bath and long process times, which would lead to the degradation of the thermally sensitive sample.[27] The solid-phase extraction (SPE) cartridge is known as one of the most used pretreatment techniques for purifying and concentrating samples. It is characterized by low cost, good repeatability and environmental friendliness compared with traditional liquid-liquid extraction. Nevertheless, to obtain size-selected polyynes from a low concentrated aqueous solution of polyynes mixture, it is not only necessary to concentrate the sample but also to separate the chains according to their length and collect each fraction.

In this work, we optimized the process parameters for the synthesis of hydrogen-capped polyynes by pulsed laser ablation in water, and we developed a single procedure entirely based on RP-HPLC able to perform multiple operations: concentration, separation, and collection of size-selected polyynes synthesized in water. Our single RP-HPLC procedure consists of the concentration performed directly on the chromatography column and a gradient method, which allows the separation and collection of the concentrated size-selected polyynes for differentiated characterizations, exploiting their structural-dependent optical properties.

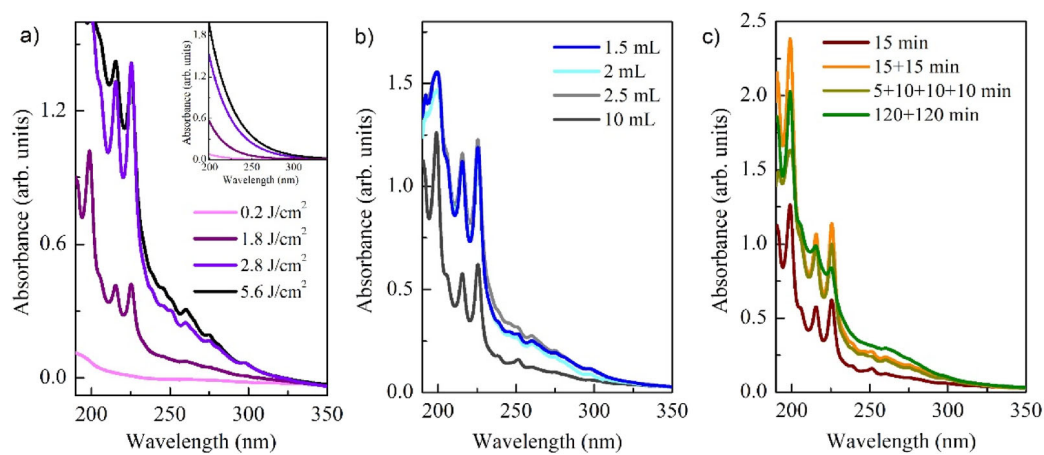

Figure 1. UV-Vis absorption spectra (range: 190–350 nm) of polyynes mixture synthesized in water: a) at different fluences (0.2, 1.8, 2.8, 5.6 J/cm$^2$), fixed water volume (2 ml) and ablation time (15 min), b) at different liquid volumes (1.5, 2, 2.5, 10 ml), fixed fluence (2.8 J/cm$^2$) and ablation time (15 min) and c) at different ablation times (15 min, 30 min with a break of 2 min after 15 min, 35 min with breaks of 2 min after 5, 15 and 25 min, 240 min with a break of 2 min after 120 min), fixed fluence (2.8 J/cm$^2$) and water volume (2 ml). In the inset, the fitting exponential curves of the background of each spectrum at different fluences (0.2, 1.8, 2.8, 5.6 J/cm$^2$).

## 2. Materials and methods

### 2.1. Chemical and reagents

A graphite target (8 mm D. × 2 mm H.) with a purity of 99.999% was acquired from Testbourne Ltd. Milli-Q water (0.055 $\mu$S) was employed as a solvent for the synthesis of polyynes by PLAL. Water as a mobile phase for RP-HPLC analysis was processed with Purelab Flex ultrapure water machine (Veolia). HPLC grade acetonitrile, employed as a mobile phase for RP-HPLC, was purchased by Aldrich.

### 2.2. Sample production

We exploited pulsed laser ablation in liquid (PLAL) for the synthesis of polyynes. Irradiation was performed by Nd:YAG nanosecond laser (Q-Smart 850, 5 ns) with second harmonic pulses (532 nm wavelength) at 10 Hz repetition rate. The laser beam was directed to the graphite target by a lens with a focal length of 200 mm. To enhance the polyynes yield, we optimized the laser power density by testing different fluences (i.e., 0.2, 1.8, 2.8, 5.6 J/cm$^2$), solvent volumes (i.e., 1.5, 2, 2.5, 10 ml) and ablation time with values ranging between 15 and 240 minutes. After laser ablation, the solution in the cell was filtered with a polytetrafluoroethylene (PTFE) syringe filter (25 mm in diameter) having a pores size of 0.45 $\mu$m. The concentration of polyynes in the mixture was estimated by UV-Vis absorbance with a Shimadzu UV-1800 UV/Visible scanning spectrophotometer (190–1100 nm).

### 2.3. RP-HPLC conditions

All separations were performed on a Shimadzu Prominence High-Performance Liquid Chromatography (HPLC) apparatus equipped with SIL-20A HT autosampler, a CTO-20A column oven, a quaternary pump LC-20AD, a SPD-M20A photodiode array UV-Vis spectrometer (DAD) and a FRC-10A fraction collector connected to LabSolutions LC/GC software. A Shimadzu Shim-pack GIST C8 column (250 mm L. x 4.6 mm I.D., 5 $\mu$m particle size) was used. A gradient system was employed for polyynes separation involving, as a mobile phase, water (A) and acetonitrile (B). The flow rate was set at 1.0 mL/min during the entire run. The column temperature was set at 40 °C and DAD operated at a sampling rate of 6.25 Hz in the range of 190–800 nm.

## 3. Results and discussion

We first report the optimization of the synthesis process parameters to reach the highest possible yield of polyynes by pulsed laser ablation in water and then we discuss our approach entirely based on RP-HPLC to concentrate, separate and collect size-selected polyynes in a single procedure.

### 3.1. Optimization of synthesis process parameters

We performed different ablations of graphite in water by changing the experimental parameters, e.g., fluence, volume of the liquid and ablation time, to optimize the polyynes formation yield, hence their concentration in aqueous solution. The spectra in Figure 1 show the absorption of polyynes produced by PLAL resulting from changing one process parameter at a time. In all the cases, we obtained a mixture of hydrogen-capped polyynes with 6, 8, and 10 carbon atoms, hereafter called $C_6$, $C_8$, and $C_{10}$, respectively.

The 0-0 bands of the $^1\Sigma_u^+ \leftarrow X^1\Sigma_g^+$ transition are at 198 nm for $C_6$, 225 nm for $C_8$, and 251 nm for $C_{10}$, as reported in the literature.[28] It is not possible to identify characteristic peaks of the longest chains since the concentration of polyynes in an aqueous solution is very low, and they also show reduced stability in water.[18,29]

The absorption spectrum of polyynes at different fluences with a fixed volume of 2 ml and ablation time of 15 minutes is shown in Figure 1a. We observe that when the fluence is too high (5.6 J/cm$^2$) or too low (0.2 J/cm$^2$), the absorption signals are not well resolved or visible at all. In the first case (i.e., high fluence), polyynes formed in the solution can be thermally degraded generating more by-products, which is proven by the higher absorption of the background (see inset of Figure 1a). Indeed, the background is associated with the concentration of by-products synthesized during PLAL simultaneously with polyynes or as a result of the degradation of sp-carbon chains.[18,20] In the second case (i.e., low fluence), the ablation of graphite and the formation of polyynes are not enough effective, hence polyynes absorption is weak or not detectable at all. Thanks to this experiment we consider 2.8 J/cm$^2$ as an optimal value for fluence, where the concentration of polyynes is even higher than the case of 1.8 J/cm.$^2$

Figure 1b shows UV-Vis absorption peaks of polyynes at different solvent volumes, for a fixed time (15 minutes) and fluence (2.8 J/cm$^2$). The highest concentration is observed in the case of 2 ml and 1.5 ml since polyynes are synthesized in less volume keeping fixed the other process parameters. Consequently, this is evidence that the crosslinking reactions between polyynes, which would lower their concentration, are not having a significant impact employing these volumes of water. We decided to select 2 ml as the optimized volume to avoid splashes which can occur at 1.5 ml.

The effect of the ablation time is shown in Figure 1c displaying the absorbance of polyynes solutions prepared at different ablation times, same volume (2 ml) and fluence (2.8 J/cm$^2$). Specifically, we ablated for: 15 minutes, 30 minutes interrupted by a break of 2 minutes after the first 15 minutes and for 240 minutes with 2 minutes break after the first 120 minutes. The breaks seem to be beneficial, allowing a decrease in the temperature of water, so polyynes are less affected by thermal degradation. To verify the effectiveness of the breaks, we performed another ablation of 35 minutes with three breaks in between: one after the first 5 minutes and the other two after every subsequent 10 minutes of ablation. The picture shows that, if the ablation time is too long, the absorbance of polyynes of different lengths decreases. Indeed, during the ablation, polyynes are continuously produced and destroyed, therefore the optimized ablation time is a matter of finding the tradeoff between the two coexisting processes.[30] Moreover, there is also an optimum for the number of breaks, since during breaks polyynes are not produced. The best result in terms of polyynes concentration is achieved when we introduced a break between the two ablations of 15 minutes, as clear from the intensity of the absorbance peaks of $C_6$, $C_8$ and $C_{10}$.

In the next paragraph, the results are related to hydrogen-capped polyynes synthesized by pulsed laser ablation in water with the optimized process parameters discussed above: 2 ml of volume, 2.8 J/cm$^2$ of fluence and 30 minutes of ablation time (with 2 minutes of break after the first 15 minutes). The concentrations of the species, computed employing the Lambert-Beer law with molar extinction coefficients taken from the literature,[31] are of the order of $10^{-6}$ mol/L for $C_6$ and $C_8$ while of the order of $10^{-7}$ mol/L for $C_{10}$.

### 3.2. Concentration, separation and collection of size-selected polyynes in water

We concentrated aqueous solutions of polyynes by RP-HPLC by adapting to sp-carbon chains the so-called batch-loading on-column concentration (i.e., on-column focusing) generally employed for protein analysis.[32,33] We effectively accumulated the polyynes mixture on the top of the column without losses by injecting 1 ml of aqueous polyynes solution while holding the chromatograph in the ready-to-run conditions at a reduced flow rate, i.e., 0.1 ml/min, and with a mobile phase of 95% water (A) and 5% acetonitrile (B). After sample loading, to separate all the polyynes, the flow rate is set to the normal value of 1 ml/min and an optimized gradient method composed of three phases is run. The first phase lasts 3 minutes, while the mobile phase is kept constant at 5% B. The second one lasts from 3 to 33 minutes during which the mobile phase linearly reaches 100% B. In the third one, the composition of the mobile phase is maintained constant at 100% B for 20 minutes, also to be sure to remove any species retained on the column. It is also possible to automatically collect each size-selected polyyne setting the times at which to trigger the fraction collector. In this way, each fraction can be extracted from the system and characterized separately. The key steps to concentrate, separate and collect size-selected polyynes starting from a low concentrated aqueous solution of polyynes mixture are summarized in Figure 2.

Each size-selected polyyne corresponds to a chromatographic peak, which in turn is related to a UV-Vis spectrum extracted from the diode array detector (DAD) coupled to RP-HPLC. Polyynes could be precisely identified by comparing UV-Vis spectra with literature data and with our theoretical simulations.[28,29,31]

In Figure 3 we reported chromatograms selected at different wavelengths (i.e., 198 nm, 225 nm, 251 nm, and 275 nm), which correspond to the maximum absorptions of different size-selected hydrogen-capped polyynes (i.e., $C_6$, $C_8$, $C_{10}$, and $C_{12}$) before and after performing our single RP-HPLC procedure. The concentration step allows us to observe an increase in the chromatographic peak intensity of $C_6$, $C_8$, and $C_{10}$ but also to detect $C_{12}$, not visible before (see the inset of Figure 3).

To quantify those increments in concentration, we calculated the ratio between the integrated area of each chromatographic peak at the wavelength of the maximum absorption of the corresponding polyyne before and after performing our single RP-HPLC procedure. Thus, we obtained the following values: 2.4 for $C_6$, 5.3 for $C_8$, and 7.7 for $C_{10}$. In the case of $C_{12}$, it was not possible to calculate this factor due to the initial hardly discernible peak. To investigate the reason why the increment of the concentration of polyynes is different according to the chain length, we focused on their UV-Vis spectra extracted from DAD of RP-HPLC and their chromatographic peaks before and after performing our single RP-HPLC procedure, as shown in Figure 4. The maximum absorption peaks of $C_6$ (198 nm) and $C_8$ (225 nm) after the concentration step appear truncated, as pointed out by the black arrows in the UV-Vis spectra reported in Figure 4a and c. In addition, the corresponding chromatographic peaks result distorted namely rounded and asymmetric (see inset of Figure 4b and d). The presence of broadened or distorted peaks of some analytes in the chromatograms is the index of an overloaded column. Hence, the increments in the concentration of $C_6$ and $C_8$ calculated from the chromatographic area extracted at these wavelengths are underestimated.[34] On the contrary, in the case of $C_{10}$, the UV-Vis spectrum (see Figure 4e) and the chromatographic peak extracted at 251 nm (see Figure 4f) are not distorted. To evaluate the actual increment of the concentration for $C_6$ and $C_8$ avoiding distortion phenomena, we chose to analyze the chromatogram extracted at different wavelengths. We selected for $C_6$ the chromatogram at 300 nm (see Figure 4b), which is related to a low-intensity absorption peak, while we extracted for $C_8$ the chromatogram at 216 nm (see Figure 4d), which is related to its second absorption peak. We calculated the ratio between the integrated area of the chromatographic peak of $C_6$ at 300 nm and $C_8$ at 216 nm before and after the concentration obtaining 8.0 and 7.5, respectively. These values are then similar to the case of $C_{10}$ (i.e., 7.7), which was not truncated.

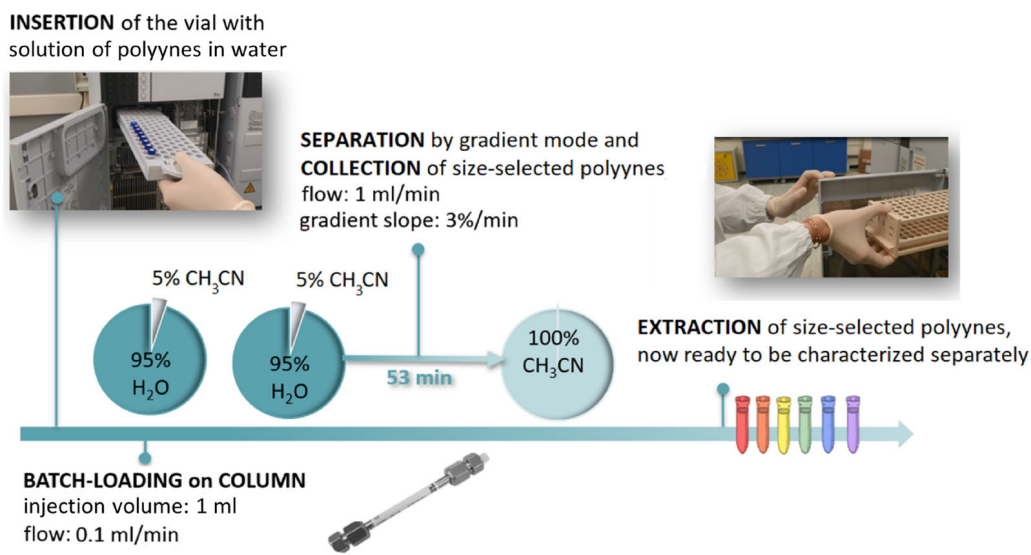

**Figure 2.** Key steps of our single RP-HPLC procedure: insertion of aqueous polyynes solution in RP-HPLC system, batch-loading on-column concentration method, separation by gradient method, collection of size-separated polyynes and extraction of the fractions.

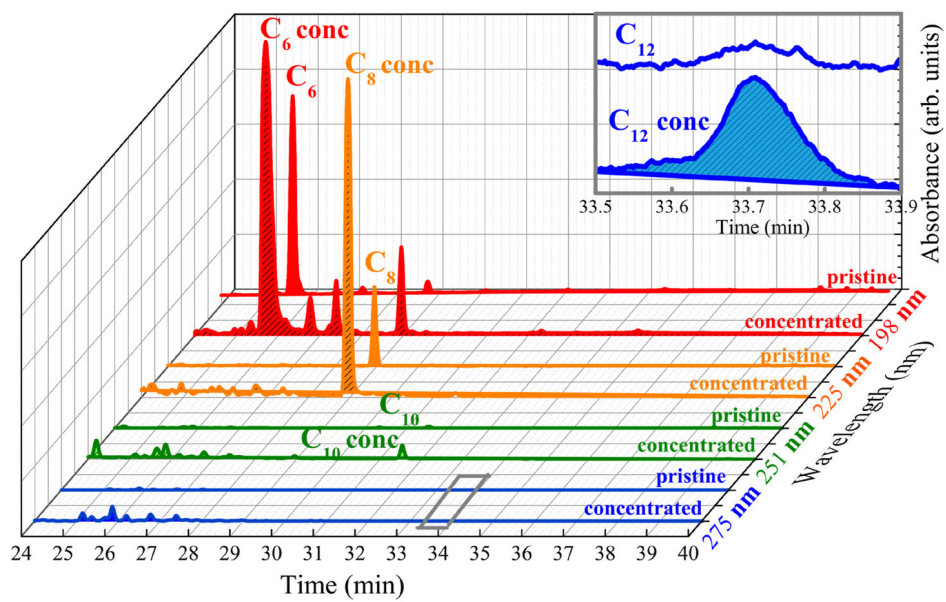

**Figure 3.** Chromatograms selected at different wavelengths to highlight the chromatographic peak of specific hydrogen-capped polyynes before and after the concentration into the RP-HPLC column (i.e., $C_6$ at 198 nm, $C_8$ at 225 nm, $C_{10}$ at 251 nm, $C_{12}$ at 275 nm). Inset: zoom on $C_{12}$ chromatographic peak before and after concentration.

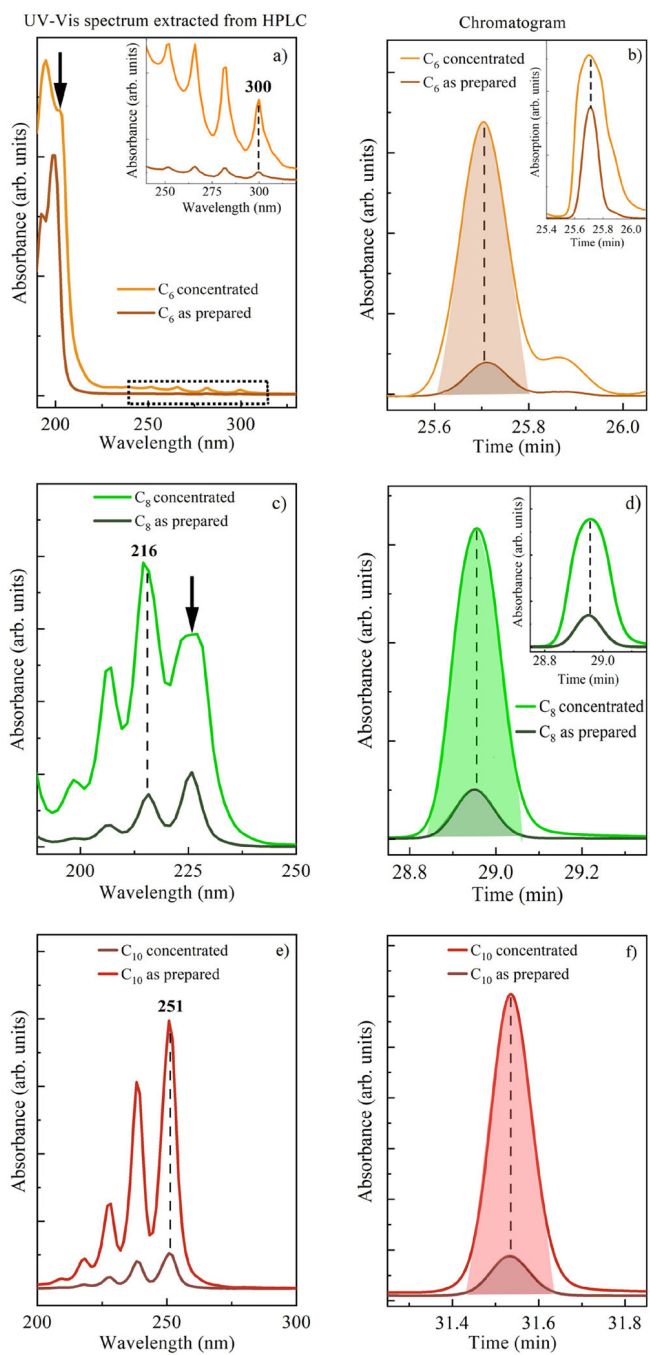

Figure 4. UV-Vis spectra (on the left column) extracted from DAD of RP-HPLC and the corresponding chromatographic peaks (on the right column) extracted at 300 nm for $C_6$, at 216 nm for $C_8$, and at 251 nm for $C_{10}$ before and after our single RP-HPLC procedure. The area considered to calculate the increments in concentration of $C_6$, $C_8$ and $C_{10}$ are colored. Insets: low-intensity absorption peaks of $C_6$ (panel a), the chromatographic peak of $C_6$ selected at 198 nm (panel b) and the chromatographic peak of $C_8$ selected at 225 nm (panel d).

In principle, we expected a concentration 10 times greater than the pristine solution since we injected in the RP-HPLC system 10 volumes of 100 μl each of the pristine sample (i.e., 1 ml of polyynes solution). Thus, we can conclude that our approach allows the increase of the concentration of polyynes synthesized in water with an efficiency up to 80 ± 3%. A possible reason that prevents our single RP-HPLC procedure to reach 100% efficiency may be related to crosslinking reactions between chains leading to polyynes degradation during the concentration step or polyyne leakage during the batch-loading on the column.

## 4. Conclusions

In this work, we employed a two-fold action to obtain higher concentrated polyynes by pulsed laser ablation in water: optimization of the synthesis process parameters to reach the highest possible yield of polyynes in an aqueous solution and the application of a novel post-synthesis procedure able not only to concentrate but also to separate and collect polyynes of different length.

The optimized parameters were identified as 2.8 J/cm$^2$ of fluence, 2 ml of solvent volume and 30 minutes of ablation time (with 2 minutes of break after the first 15 minutes), from which we detected three species of hydrogen-capped polyynes (i.e., $C_6$, $C_8$, $C_{10}$) with a concentration of $10^{-6}$-$10^{-7}$ mol/L. Then, we developed a single RP-HPLC procedure to concentrate, separate and collect size-selected polyynes. The aqueous solution of polyynes mixture needs to be inserted in the autosampler of RP-HPLC and then it will undergo our novel procedure based on two RP-HPLC methods: the first one for the concentration of the species and the second one for their separation and collection. The concentration step consists of the on-column focusing method, which is generally used for protein analysis but herein adapted for sp-carbon chains. Thanks to this, we observed the increase in the absorption of each size-selected polyyne, and we also detected another species, i.e., $C_{12}$, not visible before. After the concentration step, we noticed in the chromatographic peaks of $C_6$ at 198 nm and of $C_8$ at 225 nm and the corresponding UV-Vis spectra the distortion effects of an overloaded column which prevent the correct estimation of the increment in the concentration of $C_6$ and $C_8$. Thus, we consider the increase of the concentration of $C_6$ from the chromatographic peaks at 300 nm, $C_8$ from the chromatographic peaks at 216 nm and of $C_{10}$ from the chromatographic peaks at 251 nm to evaluate the efficiency of our single RP-HPLC procedure, which turns out to be 80 ± 3%. Furthermore, by employing columns with higher loading capacity, it could be possible to increase even more the concentration of size-selected polyynes and observe longer polyynes.

Our single RP-HPLC procedure is a useful tool for being able to fully characterize size-selected polyynes separately and test them in possible future applications. Indeed, it automizes multiple steps necessary for sample preparation as concentration, separation and collection of size-selected polyynes. Then, by successfully concentrating polyynes in water, our procedure allows obtaining results similar to the synthesis in organic solvents but with lower costs and without toxicity issues.

## Acknowledgement

The authors thank D. Giribone for his precious assistance during the development of the new procedure entirely based on HPLC able to automatize the concentration, separation, and collection of size-selected polyynes. The authors acknowledge funding from the European Research Council (ERC) under the European Union's Horizon 2020 research and innovation program ERC-Consolidator Grant (ERC CoG 2016 EspLORE grant agreement No. 724610, website: www.esplore.polimi.it).

## CRediT authorship contribution statement

**S. Peggiani:** Writing-Original Draft, Conceptualization, Methodology, Validation, Investigation, Writing-Review & Editing, Visualization.
**A Facibeni:** Conceptualization, Resources, Writing-Review & Editing.
**P. Marabotti:** Methodology, Writing-Review & Editing, Validation.
**A. Vidale:** Validation, Visualization, Writing-Review & Editing.
**S. Scotti:** Methodology, Writing-Review & Editing, Validation.
**C.S. Casari:** Conceptualization, Validation, Writing e review & editing, Supervision, Project administration, Funding acquisition.

## Disclosure statement

No potential conflict of interest was reported by the authors.

## Funding

This work was supported by European Union's Horizon 2020 research and innovation program ERC-Consolidator Grant (ERC CoG 2016 EspLORE grant agreement No. 724610, website: www.esplore.polimi.it)